\documentclass{aastex}
\usepackage{spr-astr-addons}
\usepackage{url}

\RequirePackage{color}

\begin{document}

\title{Continuum and line emission of flares on red dwarf stars} \slugcomment{Not to appear in Nonlearned J., 45.}
\shorttitle{Short article title}
\shortauthors{Morchenko et al.}

\author{E. Morchenko\altaffilmark{1}} \and \author{K. Bychkov\altaffilmark{1}}
\affil{Sternberg Astronomical Institute, Moscow M.V. Lomonosov State University, 13, University Prospect, Moscow, Russia, 119992} \and
\author{M.~Livshits\altaffilmark{2}}
\affil{N.V. Pushkov Institute of Terrestrial Magnetism, Ionosphere and Radio Wave Propagation of Russian Academy of Sciences, 4, Kaluzhskoe Hwy,
Troitsk, Moscow, Russia, 142190}

\begin{abstract}
The emission spectrum has been calculated of a homogeneous pure hydrogen layer, which parameters are typical for a flare on a red dwarf. The
ionization and excitation states  were determined by the solution of steady-state equations taking into account the continuum and all discrete
hydrogen levels. We consider the following elementary processes: electron-impact transitions, spontaneous and induced radiative transitions, and
ionization by the bremsstrahlung and recombination radiation of the layer itself. The Biberman--Holstein approximation was used to calculate the
scattering of line radiation. Asymptotic formulae for the  escape probability are obtained for a symmetric line profile taking into account the Stark
and Doppler effects. The approximation for the core of the H$-\alpha$ line by a gaussian curve has been substantiated.

The spectral intensity of the continuous spectrum, the intensity of the lines of the Balmer series and the magnitude of the Balmer jump have been
calculated. The conditions have been determined for which the Balmer jump and the emission line intensities above the continuum decrease to such low
values that the emission spectrum can be assumed to be continuum as well as the conditions at which the emission spectrum becomes close to the
blackbody.
\end{abstract}

\keywords{Red dwarfs, flares, Stark effect, H-$\alpha$ line core,
escape probability, emission spectrum}

\section{Introduction}

\cite{Gersh_72} were the first who formulated  the idea about the analogy of flares on red dwarf stars and on the Sun. At present it seems obvious
that the physical nature of the activity of the Sun and red dwarf stars is similar (\cite{Gersh_2005}).

\cite{Pik_74} put forward the pioneering idea about the important role of gas-dynamic effects in the formation of the emission of solar flares. The
authors showed that the radiative cooling of the shocked gas can be a source of the flare's emission in the optical range.
The shock wave is produced by impulsive heating of the chromosphere by suprathermal particles.

The Kostyuk--Pikel'ner model predicts the formation of a relatively
homogeneous gas layer, a few kilometers of thickness, which is
responsible for the flare's optical emission in the continuum and
spectral lines.  The layer is located between the front of the
downward shock and the region of plasma heated from above.

This model was applied to the problem of flares on red dwarf stars
in \cite{liv80}. The authors fulfilled calculations of a
one-dimensional radiative shock wave and demonstrated the formation
of a region with a small gradient of the plasma parameters:
temperature, pressure, and density (see Figures~3 and 4 in
\cite{liv80}). This result is confirmed by the calculations of the
radiative cooling behind a shock front in the atmospheres of cool
stars (see \cite{FG2001} and \cite{BBMN2014}) with a more detailed
account for the elementary processes in the plasma: ionization,
recombination, bremsstrahlung, excitation and de-excitation of
discrete levels of atoms and ions as well as radiation scattering at
the frequencies of spectral lines.

Red dwarf stars have a complex structure of flare spectra:  the intensities of the emission lines strongly increase, the continuum emission is also
enhanced; it is well approximated with Planck curves obtained for different temperatures at the blue and red sides relative to the Balmer jump (see
\cite{Kowalski13}). Layers producing the blackbody emission should be in a state close to thermodynamic equilibrium, whereas the gas generating
strong emission lines is essentially non-equilibrium.  Thus, the flare plasma includes at the same time equilibrium and non-equilibrium regions.

Therefore, we investigate the problem of the formation of an emission spectrum in a wide range of the cloud parameters. We consider a homogeneous gas
layer of thickness $\mathcal{L}$ with definite values of temperature and density and consisting of pure hydrogen.

We develop the model of \cite{dra80} in three points. Firstly, we
consider the influence of the layer's radiation (bremsstrahlung and
recombination) on the occupation of atomic levels. This development
is necessary, as the
 flare luminosity is stronger
  in the optical range  than that of
the quiescent atmosphere of the whole star. Secondly, we take into
account that broadening of the high members of spectral series is
caused exclusively by the linear Stark effect. Thirdly, we consider
the \textit{two-temperature} approximation for a flare layer.

In Section~1  the problem is formulated. In Section~2 we write the set of equations of ionization balance with regard to the continuum. In Section~3
the calculation of the continuum emission is described. In Section~4 we present the method of calculation of the quantum escape probability. In
Section~5 the intensities of the emission spectral lines are calculated. Section~6 presents the results of the emission spectrum calculations  of a
layer with parameters possible in the flare regions of red dwarf stars.

\section{Statement of the Problem}
We calculate the emission in spectral lines and continuum of an homogeneous pure hydrogen layer.  Its temperature, density and dimensions are assumed
to be given. The value of the ion--atom temperature, $T_{ai}$, is taken to be greater than the value of the electron temperature, $T_e$, because both
ions and atoms are heated more intensively than electrons at the shock front (see, e.g., \cite{Pick-1961}).

We use the results of the radiative shock wave calculations by
\cite{FG2001} and \cite{BBMN2014} to state the input parameter set.
These calculations were fulfilled under the conditions of the
atmosphere of a red giant star with an effective temperature of
$T_0=3000~K$. The values of $T_{ai}$  and $T_e$ at the shock front
are respectively $7-10$~eV and ${\approx}\,\,0.7$~eV.

The Lyman-$\alpha$ (Ly-$\alpha$) optical depth, $\tau_{12}$, in the
cooling region, as it follows from \cite{BBMN2014}, lies in the
range
\begin{equation}\label{pr_03}
1{\cdot}10^7\leqslant\tau_{12}\leqslant3{\cdot}10^7.
\end{equation}
The value of $\tau_{12}$ weakly depends on the density of the gas,
as the cooling rate, ${dS/}{dt}$\,, is proportional to the hydrogen
atom concentration, $N_a$. Thus, $N_a$ is cancelled:
\begin{equation}
d\tau_{12}\propto{}N_a{}u\cfrac{dt}{dS}\,\,dS,\,\,\cfrac{dS}{dt}\propto{}N_a\,\,\Rightarrow{}d\tau_{12}\propto{}u\,dS\,.
\end{equation}
Here,
\begin{equation}
S=T_{ai}+\cfrac{N_e}{N_\mathrm{H}}\,T_e;
\end{equation}
$N_e$ is the electron density, $u$ is the velocity of the gas
flowing from the shock front, and $dt$ is some time interval from
the moment when a given gas layer intersects the shock front.

The value of $N_\mathrm{H}$ is equal to the summa of the proton and
atom concentrations:
\begin{equation}\label{pr_04}
N_\mathrm{H}=N_p+N_a.
\end{equation}
We accept $N_\mathrm{H}$ typical for the M-dwarf chromospheres:
\begin{equation}\label{pr_05}
3{\cdot}10^{14}\text{ cm}^{-3}\leqslant{}N_{\mathrm{H}}
\leqslant3{\cdot}10^{16}\text{ cm}^{-3}.
\end{equation}
The layer thickness, $\mathcal{L}$, is defined to be such that the
condition \eqref{pr_03} is satisfied. The electron density is equal
to
\begin{equation}\label{pr_06}
N_e=N_p.
\end{equation}

The values of $T_{ai}$ and $T_e$ were obtained from a red giant
star, which has a 100x smaller gravity compared to a red dwarf star.
The ion-atom temperature at the shock front is determined by the
shock velocity, $u_0$, and weakly depends on the density of the gas.
If the value of $u_0$ lies in the range $50{-}70$ km/sec, then it is
true that
\begin{equation}\label{pr_01}
6\text{ eV}\leqslant{}T_{ai}\leqslant12\text{ eV}.
\end{equation}
The value of $T_e$ at the shock front is determined by the
temperature $T_0$ of the unperturbed gas and heating due to the
compression. The evolution of $T_e$ is caused by the balance between
the supply of energy in elastic collisions with ions and atoms and
radiative losses in the Lyman-$\alpha$ line. The radiative cooling
in the hydrogen lines is the most intensive at $T_e\sim1\text{ eV}$
(see \cite{BBMN2014}). Therefore, we assume that
\begin{equation}
0.8\text{ eV}\leqslant{}T_{e}\leqslant1.5\text{ eV}.
\end{equation}

\section{Balance Equations for the Populations of the Levels}

The following processes we take into account: the electron impact ionization, excitation, and de-excitation, the triple recombination, radiative
recombination (induced and spontaneous) as well as the influence of the layer's  bremsstrahlung and recombination radiation on its ionization and
excitation states.

The impact coefficients are taken from \cite{Johnson72}. We use the Kramers approximation (see, e.g., \cite{sob63}) for the photoionization cross
sections from the excited levels. The rates of the inverse processes are calculated following the detailed balance principle.  The spectral lines
optical depth can be rather large in this problem, so the scattering of line radiation should be taken into account. We introduce so called effective
Einstein transition coefficients for  the transitions between hydrogen levels $i$ and $k$ by the following formulae:
\begin{equation}\label{b_00}
\left\{
\begin{array}{c}
  A^*_{ki}\\
  B^*_{ki}\\
  B^*_{ik} \\
\end{array}
\right\}= \left\{
\begin{array}{c}
  A_{ki}\\
  B_{ki}\\
  B_{ik} \\
\end{array}
\right\}\times\theta_{ik}.
\end{equation}
Here, $\theta_{ik}$ is a photon escape probability out of the process of scattering. The values of $\theta_{ik}$ are calculated in the section
\ref{probab} for the resonance as well as subordinate transitions.

According to the standard notation, $A_{ki}$ and $B_{ki}$ are respectively the probabilities of spontaneous and induced transition from the upper level
$k$ to the lower level $i$, and $B_{ik}$ is the absorption coefficient.

We write the set of the balance equations for the discrete levels of a hydrogen atom:
\begin{subequations}\label{b_01}
\begin{align}
\sum\limits_{j=1}^{k-1}a_{1j}N_j&+a_{1k}N_k+\sum\limits_{j=k+1}^{n}a_{1j}N_j=%
\alpha_1N_eN_p,\label{b_01a} \\
\sum\limits_{j=1}^{k-1}a_{2j}N_j&+a_{2k}N_k+\sum\limits_{j=k+1}^{n}a_{2j}N_j=%
\alpha_2N_eN_p,\label{b_01b} \\
&\hbox to 2cm{\dotfill} \notag \\
\sum\limits_{j=1}^{k-1}a_{kj}N_j&+a_{kk}N_k+\sum\limits_{j=k+1}^{n}a_{kj}N_j=%
\alpha_kN_eN_p,\label{b_01c} \\
&\hbox to 2cm{\dotfill} \notag \\
\sum\limits_{j=1}^{k-1}a_{nj}N_j&+a_{nk}N_k+\sum\limits_{j=k+1}^{n}a_{nj}N_j=%
\alpha_nN_eN_p,\label{b_01d}
\end{align}
\end{subequations}
where
\begin{equation}
\alpha_i={}r_i+\gamma_iN_e,\qquad{}i=1,\ldots{},n.
\end{equation}
We assume that the maximum value of the principal quantum number, $n$, is equal to an effective series limit. It can be evaluated by the Inglis--Teller
formula (\cite{vid66}), as
\begin{equation}\label{b_02}
\lg{}n=3.10-0.13\lg(2N_e/{\mathrm{cm}^{-3}}).
\end{equation}
The following designations are introduced:\\
$N_k$ --- the density number of the hydrogen atoms on the discrete level
with the principal quantum number $k$;\\
$r_k,\gamma_k$ --- the radiative and triple recombination coefficients on the $k$-th level, respectively;\\
$a_{ik}\;(i\neq{}k)$ --- the transition rate from the level $k$ to the level
$i$;\\
$a_{kk}$ --- the escape rate from the level  $k$.

We write the explicit formulae for the matrix coefficients in
the left side of the equation set.\\
Excitation rate ($k{<}i$) is:
\begin{equation}\label{b_03}
a_{ik}=-\left[q_{ki}(T_e)N_e+4\pi{}I_{ki}{}B^*_{ki}\right],
\end{equation}
where\\
$q_{ki}(T_e)$ is the electron impact excitation;\\
$I_{ki}$ is the spectral intensity of the layer continuous radiation
in the frequency of $k{\leftrightarrow}i$ transition.\\
Deactivation rate ($k{>}i$) is:
\begin{equation}\label{b_05}
a_{ik}=-\left[q_{ki}(T_e)N_e+A^*_{ki}+4\pi{}I_{ki}B^*_{ki}\right],
\end{equation}
where $q_{ki}(T_e)$ is the impact deactivation coefficient.

The diagonal elements are
\begin{multline}\label{b_07}
a_{kk}=\sum_{j=1}^{k-1}\left[q_{kj}(T_e)N_e+A^*_{kj}+%
4\pi{}I_{kj}B^*_{kj}\right]+\\%
+\sum_{j=k+1}^{n}\left[q_{kj}(T_e)N_e+%
4\pi{}I_{kj}B^*_{kj}\right]+\mathcal{P}_k+q_k\cdot{}N_e,
\end{multline}
where $q_k$ is the electron impact ionization coefficient, and $\mathcal{P}_k$ is the photoionization rate from the level $k$:
\begin{equation}\label{b_08}
\mathcal{P}_k=4\pi\int_{\nu_k}^\infty\dfrac{I_\nu}{h\nu}%
\,\sigma_k(\nu)\,d\nu\,.
\end{equation}
Here, $\nu_k$ is the photoelectric threshold from level $k$,  $h$ is the Plank constant, $\sigma_k(\nu)$ is the cross section in the Kramers
 approximation:
\begin{equation}\label{b_09}
\sigma_k(\nu)=%
\dfrac{64}{3\sqrt3}\,\alpha\pi{}a_0^2\cdot{}k^{-5}%
\cdot\left(\dfrac{\mathrm{Ry}}{h\nu}\right)^3\,,
\end{equation}
where $\alpha$ is the fine structure constant, $a_0$ is the Bohr radius.\\
The spontaneous recombination coefficient, $r_k^{(s)}$, can be calculated from \eqref{b_09} using the principle of detailed balance:
\begin{subequations}\label{b_10}
\begin{align}
r_k^{(s)}&=\mathcal{R}(\beta_k)%
E_1(\beta_k),\qquad\beta_k=\cfrac{\mathrm{Ry}}{k^2T_e}\,;\label{b_10a}\\
\mathcal{R}(x)&=C_r\cdot{}x^{3/2}e^x\,,\qquad{}C_r=\dfrac{2^6}{3}\sqrt{\dfrac{\pi}3}\,\alpha^4c{}a_0^2\,,\label{b_10b}
\end{align}
\end{subequations}
where $E_1(x)$ is the exponential integral of order 1.

The coefficients of induced recombination, $r_k^{\mathrm(\mathrm{ind})}$, is given in the next section. It is true that
\begin{equation}\label{s_15}
r_k=r_k^{(s)}+r_k^{(\mathrm{ind})}.
\end{equation}
Summing  the set \eqref{b_01} over all discrete levels yields  the identity
\begin{equation}\label{b_11}
\sum_i\left(\mathcal{P}_i+q_iN_e\right)\cdot{}N_i=%
\sum_i\left(r_i+\gamma_iN_e\right)\cdot{}N_eN_p.
\end{equation}
As a result, we come to the  set of homogeneous equations with singular matrix for the relative level occupations, $y_i{=}N_i/N_a$:
\begin{equation}\label{b_12}
\sum_{j=1}^n\tilde{a}_{ij}y_j=0,\,\,i=1,\ldots{},n.
\end{equation}
where
\begin{align}
\tilde{a}_{ik}&=a_{ik}-%
\dfrac{\alpha_i}{\Upsilon}\left(\mathcal{P}_k+%
q_kN_e\right),\label{b_14}\\
\alpha_i&=r_i+\gamma_iN_e,%
\qquad{}\Upsilon=\sum_{j=1}^n\alpha_j.\label{b_15}
\end{align}
To avoid the singularity, the last equation
 ($k{=}n$) is replaced
by the normalization condition
\begin{equation}\label{b_13}
\sum_{k=1}^ny_k=1.
\end{equation}
We take into account the scattering of the ionizing radiation using the division of $r_k$ by $1+\tau_k$, where $\tau_k$ is the optical depth on the
photoionization threshold from the level $k$.

\section{Continuum Emission}
Continuum emission consists of two components: free-free and bound-free. The following formula is valid for an homogeneous layer:
\begin{equation}\label{s_01}
I_\nu=S_\nu\cdot\left(1-e^{-\tau_\nu}\right),
\end{equation}
where $S_\nu$ is the source function, which is equal to the ratio of the emission coefficient, $\varepsilon_\nu$, to the absorption coefficient,
$\varkappa_\nu$:
\begin{equation}\label{s_02}
S_\nu=\dfrac{\varepsilon_\nu}{\varkappa_\nu}=%
\dfrac{\varepsilon^{(b)}_\nu+%
\sum\limits_{j=k}^n\varepsilon^{(r)}_{\nu{}j}}%
{\varkappa^{(b)}_\nu+%
\sum\limits_{j=k}^n\varkappa^{(i)}_{\nu{}j}}.
\end{equation}
The summation is over all the levels for which the frequency of the photoelectric threshold is less than  $\nu$:
\begin{equation}\label{s_02a}
h\nu\geqslant{}\cfrac{\mathrm{Ry}}{k^2}\,.
\end{equation}
The Menzel factor is designated by $b_k$:
\begin{equation}\label{s_03}
b_k=\dfrac{N_k}{N_eN_p}\,\dfrac{\left(2\pi{}m_eT_e\right)^{3/2}}%
{k^2h^3}\,%
\exp({-\beta_k}).
\end{equation}
It shows the deviation of the level population from its equilibrium value. This factor is equal to 1 when excitation and ionization are described by
the formulae of Boltzmann and Saha.

We use the Kramers approximation for the absorption coefficients (\cite{sob63}, \textsection{}34.4):
\begin{subequations}\label{s_05}
\begin{align}
\varkappa^{(b)}_\nu&=\dfrac{2^8}{3\sqrt{3}}%
\alpha\,\pi^{5/2}a_0^5%
\sqrt{\dfrac{\mathrm{Ry}}{T_e}}%
\left(\dfrac{\mathrm{Ry}}{h\nu}\right)^3
\left(1-e^{-y}\right)N_eN_p,\label{s_05a} \\[1ex]
\varkappa^{(i)}_{\nu{k}}&%
=\dfrac{2^{9}}{3\sqrt{3}}%
\alpha\,\pi^{5/2}a_0^5\dfrac{1}{k^3}b_k%
\left(\dfrac{\mathrm{Ry}}{h\nu}\right)^3%
\left(\dfrac{\mathrm{Ry}}{T_e}\right)^{3/2}\times\notag\\%
&\quad\times\left(1-\dfrac1{b_k}\,e^{-y}\right) e^{\beta_k}N_eN_p,\qquad{}y\geqslant\beta_k,\label{s_05b}
\end{align}
\end{subequations}
where $y=\dfrac{h\nu}{T_e}$\,.

We write the relation between the ionization coefficient from the level $k$, $\varkappa^{(i)}_{\nu{}k}$,  and the bremsstrahlung coefficient,
$\varkappa^{(b)}_\nu$:
\begin{equation}\label{s_08}
\dfrac{\varkappa^{(i)}_{\nu{}k}}{\varkappa^{(b)}_\nu}%
=\varphi_k\dfrac{b_k-e^{-y}}{1-e^{-y}},\qquad\varphi_k\equiv%
\dfrac{2}{k}\beta_ke^{\beta_k}.
\end{equation}
The value of $\tau_\nu$ is equal to the sum of  bremsstrahlung optical depth, $\tau^{(b)}_\nu$, and photoionization one, $\tau^{(i)}_\nu$:
\begin{equation}\label{s_09}
\tau_\nu=\tau^{(b)}_\nu+\tau^{(i)}_\nu=%
\left(\varkappa^{(b)}_\nu+\sum_{j=k}^n\varkappa^{(i)}_{\nu{}j}\right)\cdot\mathcal{L}.
\end{equation}
The free-free absorption and emission coefficients are related by Kirchhoff low:
\begin{equation}\label{s_10}
\dfrac{\varepsilon^{(b)}_\nu}{\varkappa^{(b)}_\nu}=B_\nu(T_e).
\end{equation}
The Menzel factor connects the radiative recombination coefficient, $\varepsilon^{(r)}_{\nu{}k}$, and photoionization coefficients for the $k$-th
level:
\begin{equation}\label{s_11}
\dfrac{\varepsilon^{(r)}_{\nu{}k}}{\varkappa^{(i)}_{\nu{k}}}=%
B_\nu(T_e)\cdot%
\dfrac{1-e^{-y}}{b_k-e^{-y}}\,,\qquad\left(y\geqslant\beta_k\right).
\end{equation}
Using the last formula and taking into account \eqref{s_02a}, we can transform \eqref{s_02}:
\begin{equation*}
S_\nu=\dfrac{\varepsilon^{(b)}_\nu/\varkappa^{(b)}_\nu+%
\sum\limits_{j=k}^n\left(\varepsilon^{(r)}_{\nu{}j}/%
\varkappa^{(i)}_{\nu{}j}\right)%
\cdot\left(\varkappa^{(i)}_{\nu{}j}/\varkappa^{(b)}_\nu\right)}%
{1+\sum\limits_{j=k}^n%
\varkappa^{(i)}_{\nu{}j}/\varkappa^{(b)}_\nu}=
\end{equation*}
\begin{equation}\label{s_12}
=B_\nu(T_e)\cdot\dfrac{1+\sum\limits_{j=k}^n%
\dfrac{\varkappa^{(i)}_{\nu{}j}}{\varkappa^{(b)}_\nu}\cdot%
\dfrac{1-e^{-y}}{b_j-e^{-y}}}%
{1+\sum\limits_{j=k}^n\dfrac{\varkappa^{(i)}_{\nu{}j}}%
{\varkappa^{(b)}_\nu}}=%
B_\nu(T_e)\cdot\psi_k(y),
\end{equation}
where the following function $\psi_k$ is introduced:
\begin{equation}\label{s_13}
\psi_k(y)\equiv\dfrac{1+\sum\limits_{j=k}^n\varphi_j}%
{1+\dfrac1{1-e^{-y}}\sum\limits_{j=k}^n\varphi_j\cdot(b_j-e^{-y})}.
\end{equation}
The induced recombination coefficient is equal to
\begin{equation}\label{s_14}
r_k^{(\mathrm{ind})}=\mathcal{R}(\beta_k)\cdot%
\int_{\beta_k}^\infty%
\dfrac{\psi_k(y)\cdot\left(1-e^{-\tau_\nu}\right)}%
{e^y-1}\,\dfrac{e^{-y}}{y}\,dy.
\end{equation}
As is easily seen from \eqref{s_05}, the continuum optical depth, $\tau_\nu$, is the function of $y$ when the frequency is varied and the other
parameters are unchanged.

\section{Photon escape probability}\label{probab}
The source of optical emission moves slowly and cohesively, i.e. there is no noticeable velocity gradient. Therefore, the Biberman-Holstein
approximation ({\cite{hol47}}) is used to calculate the scattering of line radiation. Consider a stationary gas that occupies a volume $V$ bounded by
a closed surface $S$. Let $\mathbf{r}$ is the radius-vector of a point within the volume. According to \citet{bib82}, the photon escape probability
from this point outside the homogeneous plasma can be written in the form:
\begin{equation*}\label{Bib_2}
\theta(\mathbf{r})=\int{}d\omega\,\,a_\omega{}%
\oint\limits_{S}\cfrac{(\mathbf{dS},%
\mathbf{R}-\mathbf{r})}{4\pi|\mathbf{R}-\mathbf{r}|^3} \exp\left({-k_{\omega}|\mathbf{R}-\mathbf{r}|}\right).
\end{equation*}
Here, $\omega$ is the angular frequency, $a_\omega$ is the spectral line profile, and
 $k_\omega$ is the line absorption coefficient. $\mathbf{R}$ denotes the radius-vector of a point belonging to the surface $S$.

Let the gas occupies a layer of finite thickness $\mathcal{L}$. In this case, the mean escape probability is equal to
\begin{equation}\label{probability}
{\theta}=\int{}d\omega\,\,a_\omega{}\mathcal G(\tau_\omega)\,,
\end{equation}
where
\begin{equation}
\mathcal G(\tau_\omega)=\cfrac{1}{4\tau_\omega}\,[1-2E_3(2\tau_\omega)]\,,\,\,\,\tau_\omega=\cfrac{k_\omega{}\mathcal{L}}{2}\,\,.
\end{equation}
In these expressions, $\tau_\omega$ is the optical depth at the center of the layer and $E_3(x)$ is the third-order exponential integral (see, e.g.,
 \cite {smi72}).

 As stated in \cite{sob96}, the very wide wings of the Balmer lines in the optically {\itshape thick} flare plasma
are due to the linear Stark effect. When the Stark broadening by electrons and ions is treated in a
 semi-classical picture, the static limit is valid in the distant long-wavelength line wing (\cite{sob63}, \textsection38.1):
 \begin{equation}\label{st_lim}
 a_\omega^{\mathrm{St}}
 {}d\omega\approx%
 \cfrac{15}{4\sqrt{2\pi}}\cfrac{({\mathcal{B}_{kn'}%
 \mathcal{E}_{0}})^{3/2}}{\Delta{\omega}^{5/2}_{kn'}}\,d\omega\,.
 \end{equation}
Here, ${\mathcal{B}}_{kn'}$ is the Stark broadening parameter of the transition from level $k$ to level $n'$ ($k>n'$), $\mathcal{E}_{0}$ is the
normal Holtsmark field, and $\Delta\omega_{kn'}=\omega-\omega_{kn'}$, \,$\omega_{kn'}$ is the transition angular frequency.  The Stark width is
approximately equal to
\begin{equation}\label{Stark_eff}
\mathcal{B}_{kn'}\mathcal{E}_{0}=%
\left(\cfrac{\pi}{2}\right)^{2/3}\,%
\cfrac{\hbar}{m_e}\,(k^2-{n'}^2)\,N_p^{2/3}\,.
\end{equation}

The spectral line profile is determined by radiation damping, Doppler effect and pressure effects. The extensive tables of the Stark broadened
profiles for hydrogen plasma at $T_{ai}=T_e$ have been published by \cite{ste94}. In the case of $T_{ai}>T_e$ we will use a simple model of the line
profile with the Doppler core and the Holtsmark wings.

We consider, e.g., the H-$\alpha$ line profile in order to justify this model. The following set of parameters is assumed:
\begin{equation}\label{Par3}
 N_e=10^{14}\,\mbox{cm}^{-3}\,,\,\,\,T_{ai}=6\,\mbox{eV}\,,\,\,\,T_e=1\,\mbox{eV}\,.
\end{equation}
We introduce the value of $\Omega_{32}$ such that
\begin{equation*}
\Omega_{32}={v^2}\,\left[6\cfrac{e^2a_0}{\hbar}\right]^{-1}\,,\,\,\mbox{where}\,\,v\,\,\mbox{is the hydrogen atom}
\end{equation*}
velocity (along a line of sight). The value of $v$ can be evaluated as
\begin{equation}
v\sim\sqrt{\cfrac{3T_{ai}}{m_H}}\,\,\Rightarrow\,\,\Omega_{32}\sim2.87\cdot10^{12}\,\,\mbox{sec}^{-1}.
\end{equation}
Here, $m_{H}$ is the hydrogen atom mass. The Doppler width, $\Delta{\omega_{kn'}^D}$, is equal to
\begin{equation}\label{Dopler}
\Delta{\omega_{kn'}^D}=\omega_{kn'}\sqrt{\cfrac{2T_{ai}}{m_{H}c^2}}\,.
\end{equation}
Here, $k=3$, $n'=2$ $\Rightarrow$ $\Delta{\omega_{32}^D}\approx3.246\cdot10^{11}\,\,\mbox{sec}^{-1}$ and $\Delta{\omega_{32}^D}\ll\Omega_{32}$.
Therefore, the {\itshape static limit} (\ref{st_lim}) can be used when $\Delta{\omega}_{32}\gg\Omega_{32}$, i.e., in the line wing (see \cite{sob63},
\textsection38.1, \textsection36.6).

 On the other hand, we introduce the parameter $\gamma$, which is equal to the sum of the radiative damping constant, $\gamma_{\mathrm{rad}}$, and the parameter $\gamma_2$:
\begin{equation}
\gamma=\gamma_{\mathrm{rad}}+\gamma_2;\,\,\gamma_2\sim{72\pi}^3N\,v^{-1}\sim5.4\cdot10^{-4}N,
\end{equation}
where $N$ is the concentration of perturbing particles.   $\gamma_{\mathrm{rad}}<2\cdot4.67\cdot10^{8}$ sec$^{-1}$\,\,$\Rightarrow$
$\gamma\sim\gamma_2$. If $N\sim{}N_e$ then ${\gamma}{/(\Delta{\omega_{32}^D})}\ll1$. Therefore, the central part of the H-$\alpha$ line profile is
the {\itshape{}Doppler profile} (see \cite{sob63}, \textsection36.6).

 In general, there is some analogy between the line profile with the Doppler core and the Holtsmark wings  and the Voigt profile when the Voigt parameter $a$
 is much less  than 1.
The normalized Voigt function is
\begin{equation}\label{Voigt_f}
a_{\omega}^V=\cfrac{1}{\sqrt{\pi}\Delta{\omega_{kn'}^D}}\,H(a,v)\,,
\end{equation}
where  $v$ is the dimensionless frequency displacement from the line center in units of  $\Delta{\omega_{kn'}^D}$.
For $a\ll1$ it can be assumed that
in the first approximation (see, e.g., \cite{iva69})
\begin{equation}\label{Voigt_as}
H(a,v)=%
\begin{cases}%
\exp{(-v^2)}\,,& \,\,0\leq|v|\leq|v_0|\\
\cfrac{a}{\sqrt{\pi}v^2}\,,& \,\,|v|\geq|v_0|\,.\\
\end{cases}\\[1ex]
\end{equation}
 The value of $|v_0|$ is the solution of the equation
\begin{equation}\label{Voigt_v0}
\exp{(-v_0^2)}=\cfrac{a}{\sqrt{\pi}v_0^2}\,.
\end{equation}
The transition from the Doppler core to the Lorentz wings in the Voigt profile is quite sharp (\cite{mih78})\,.

\subsection{Ly-$\alpha$ line}
The following model {\itshape symmetric} profile is similar to the Voigt profile (\ref{Voigt_f}):
\begin{equation}\label{St_f}
a_\omega{}=\cfrac{C}{\sqrt{\pi}\Delta{\omega_{kn'}^D}}\,\Phi(b_{kn'},v)\,,
\end{equation}
where $C$ is a normalizing constant. The function $\Phi$ is defined by
\begin{equation*}\label{Stark_as}
\Phi(b_{kn'},v)=%
\begin{cases}%
\exp{(-v^2)}\,,& \,\,0\leq|v|\leq|v_0|\\
\cfrac{15}{4\sqrt{2}}\,\cdot\cfrac{b_{kn'}}{v^{5/2}}\,,& \,\,|v|\geq|v_0|\,.\\
\end{cases}\\[1ex]
\end{equation*}
In these expressions, the value of parameter $b_{kn'}$ is significantly less than unity:
\begin{equation}\label{b_par}
b_{kn'}\equiv\left(\cfrac{\mathcal{B}_{kn'}\mathcal{E}_{0}}{\Delta\omega_{kn'}^D}\right)^{3/2}\ll1\,.
\end{equation}
We can find the constant $C$ from the equations
\begin{equation}\label{Eq4f}
1=\int\limits_{-\infty}^{+\infty}a_\omega{}\,d\omega=C\cdot\left(\mathrm{erf}\,{v_0}+\cfrac{5}{\sqrt{2\pi}}\,\cdot\cfrac{b_{kn'}}{v_0^{3/2}}\right)\,,
\end{equation}
\begin{equation}\label{Eq_4}
\cfrac{15}{4\sqrt{2}}\,\cdot\cfrac{b_{kn'}}{v_0^{5/2}}=\exp(-v_0^2)\,.
\end{equation}
  \cite{dra80} introduced the model profile phenomenologically. Here, we give the theoretical foundation for its application.

To make an estimate of the optical depth in the Ly-$\alpha$ line at $|v|=|v_0|$, we use (\ref{Par3}). According to (\ref{Stark_eff}), the Stark width
of the Ly-$\alpha$ line is approximately equal to $\mathcal{B}_{21}\mathcal{E}_{0}=1,01\cdot10^{10}\,\,\mbox{sec}^{-1}$\,, the Doppler width is
$\Delta{\omega_{21}^D}=1.75\cdot10^{12}\,\,\mbox{sec}^{-1}$\,; whence it follows that
\begin{equation*}
b_{21}=4.385\cdot10^{-4}\ll1\,.
\end{equation*}
Thus, the approximation (\ref{St_f}) is adequate in our problem. Using (\ref{Eq4f}) and (\ref{Eq_4}), we find that $|v_0|\approx3.095$ and
 $C=1$\,. If the optical depth at the line center is equal to $\tau_{\omega_0}^D=3\cdot10^7$, then it is true that
\begin{equation*}
\tau_{\omega}(|v_0|)=\tau_{\omega_0}^D\,\cdot\exp(-v_0^2)=2\cdot10^3\gg1\,.
\end{equation*}
Therefore, the Ly-$\alpha$ photon lives the medium at the frequencies of the distant line wings.

As a result, equation (\ref{probability}) can be written in the form:
\begin{equation}\label{Eq_1}
\theta_{12}=b_{21}\cfrac{15}{\sqrt{8\pi}}\,\int\limits_0^{+\infty}\mathcal
G(\tau_\omega)\cfrac{dv}{v^{5/2}}\,;\,\,\,\,\tau_\omega=\tau_\omega({v})\,.
\end{equation}
It is obvious that
\begin{equation*}
k_\omega=k_{\omega_0}^D\,\cfrac{15}{4\sqrt2}\,\cdot\cfrac{b_{21}}{v^{5/2}}\,,\,\,\,\,\,\,\tau_\omega=\tau_{\omega_0}^D\,\cfrac{15}{4\sqrt2}\,\cdot\cfrac{b_{21}}{v^{5/2}}\,.
\end{equation*}
Here,  $k_{\omega_0}^D$ is the absorption coefficient at the Doppler profile center:
\begin{equation}
k_{\omega_0}^D=\cfrac{2\pi^2e^2}{m_ec}\,f_{n'k}N_{n'}\cfrac{1}{\sqrt{\pi}\Delta{\omega_{kn'}^D}}\,;\,\,\,\,\tau_{\omega_0}^D=\cfrac{{k_{\omega_0}^D}{}\mathcal{L}}{2}\,,
\end{equation}
 $f_{n'k}$ is the oscillator strength for absorption ($n'=1$, $k=2$). Taking into account that $\tau_{\omega_0}^D\gg1$, we obtain the asymptotic
formula for the mean escape probability:
\begin{equation*}
\theta_{12}=\cfrac{4}{5}\,\cdot\left(\cfrac{15}{4\sqrt{2}}\right)^{2/5}\,\cfrac{b_{21}^{2/5}}{\sqrt{\pi{}}\,{(\tau_{\omega_0}^D)}^{3/5}}\,
\int\limits_0^{\tau_{\omega_0}^D\rightarrow+\infty}{\mathcal G(\tau_{\omega})}\,\cfrac{d\tau_\omega}{\tau_{\omega}^{2/5}}\,.
\end{equation*}
Consider the integral
\begin{equation*}\label{Eq_2}
I=\int\limits_0^{+\infty}{\mathcal
G(\tau)}\,\cfrac{d\tau}{\tau^{\kappa}}=\cfrac{1}{4}\int\limits_0^{+\infty}[1-2E_3(2\tau)]\,\cfrac{d\tau}{\tau^{1+\kappa}}\,,\,\,\,\kappa\in(0,1)\,.
\end{equation*}
By using the formula of integration by parts, this equation can be represented in the form:
\begin{equation*}\label{Eq_3}
I-\cfrac{1}{\kappa}\int\limits_0^{+\infty}E_2(2\tau)\,\cfrac{d\tau}{\tau^{\kappa}}=-\cfrac{1}{4\kappa}\left.\cfrac{\left[1-2E_3(2\tau)\right]}{\tau^\kappa}\right|_0^{+\infty}\,.
\end{equation*}
If the value of $\kappa$ lies between 0{} and 1{}, then the r.h.s. of the last equation tends to zero. Thus, we obtain that
\begin{equation}
I=\cfrac{\mu^2}{2\mu-1}\,\cfrac{\Gamma(1-{1}/{\mu})}{2^{1-{1}/{\mu}}}\,,\,\,\,\,\mu\equiv{\kappa}^{-1}\,,
\end{equation}\\
where $\Gamma$ is the gamma function. In this problem, the value of $\mu$ is equal to $5/2$. Finally, we come to
\begin{equation}\label{Stark}
\theta_{12}\approx\left(\cfrac{\mathcal{B}_{21}\mathcal{E}_{0}}{\Delta\omega_{21}^D}\right)^{3/5}\,\cfrac{1}{({\tau_{\omega_0}^D})^{3/5}}\,.
\end{equation}
The exponent $3/5$ in the last equation is the feature of the linear Stark effect.

\subsection{Generalization for the case of a multilevel atom}
Strictly speaking, the resulting expression for $\theta_{12}$ is valid only for a model two-level atom. However, the escape probability approach can
be applied also for a multilevel atom. We will use  the model profile (\ref{St_f}) up to $\bar{b}=3\cdot10^{-2}$ (for more details, see
\cite{ste94}). Then the principal quantum numbers, $k$ and $n'$, must satisfy the condition
\begin{equation}\label{Cond}
\bar{b}\geqslant%
b_{kn'}=\left(\cfrac{{k}^2{{n'}}^2}{4}\right)^{3/2}\,b_{21}\,.
\end{equation}
Constant $C$ is approximately unity in the range $10^{-4}\leqslant{}b_{kn'}\leqslant10^{-2}$; therefore, it is true that
\begin{equation}\label{Stark_itog1}
\theta_{n'k}\approx\left(\cfrac{\mathcal{B}_{kn'}\mathcal{E}_{0}}%
{\Delta\omega_{kn'}^D}\right)^{3/5}\,\cfrac{1}{({\tau_{\omega_0}^D})^{3/5}}\,,
\end{equation}
where $\tau_{\omega_0}^D$ is the optical depth in the $n'\rightarrow{}k$ transition\,.

 Since the Stark width increases with increasing $k$, the Stark broadening dominates over the Doppler broadening in the limit $k\gg1$. Then we
 use the following model profile:
\begin{equation}\label{Griem}
a_{\beta}=%
\begin{cases}%
a_0\,,& \,\,0\leq|\beta|\leq|\beta_0|\\
\cfrac{3}{2}\,\cfrac{1}{\beta^{5/2}}\,,& \,\,|\beta|\geq|\beta_0|\,,
\end{cases}\\[1ex]
\end{equation}
Here, $a_0$ is some constant, $\beta\equiv\cfrac{\omega-\omega_0}{\mathcal{B}_{kn'}\mathcal{E}_{0}}$\,.
  The normalization condition gives
\begin{equation*}
\beta_0\approx3\,,\,\,\,a_{\omega_0}\approx\cfrac{0.1}{\mathcal{B}_{kn'}\mathcal{E}_{0}}\,.
\end{equation*}
After some calculations we obtain that
\begin{equation}\label{Stark_2}
\theta_{n'k}\approx\cfrac{0.36}{\tau_{\omega_0}^{3/5}}\,,\,\,\tau_{\omega_0}=\cfrac{\pi^2e^2}{m_ec}\,f_{n'k}N_{n'}a_{\omega_0}\mathcal{L}\,.
\end{equation}

\subsection{Application to the observations of the H-${\alpha}$ line profile}
It is interesting to note that the profile with the Doppler core and the Stark-broadened wings is similar
 to the H-${\alpha}$ line profile in the flare on UV Ceti (dM5.6e) published in
\cite{eas92}. The authors found a Doppler profile  to fit the line {\itshape core}: for H-${\alpha}$ during outburst, the Doppler width was equal to
0.9 \AA{}. Thus, it is true that $T_{ai}\approx8.8$ eV and $\Delta\omega_{32}^{D}\approx3.93\cdot10^{11}\,\,\mbox{sec}^{-1}$. Further, Eason et al.
found that $\mathrm{log}({N_e})=14.75$ $\Rightarrow$ $B_{32}\mathcal{E}_{0}\approx5.33\cdot10^{10}\,\,\mbox{sec}^{-1}$ and
$\gamma\sim2.5\cdot10^{11}\,\,\mbox{sec}^{-1}$. Thus, we obtain that ${\gamma}/{\Delta\omega_{32}^{D}}\sim0.6$ and $b_{32}\ll1$.

The authors satisfactorily represented {\itshape part} of the long-wavelength wing with a Stark profile from the paper by \cite{und59}. In this paper
the Stark profile was calculated taking into account only the ion broadening in context of the Holtsmark distribution of the electric microfields,
while the Stark broadening by electrons was neglected. But as stated in {\cite{sob95}}, the electron broadening dominates in stellar flares. We
consider that the deviation of the observed H$\alpha$ line profile from the Stark profile in the red wing (see Fig. 9 in \cite{eas92}) is almost
entirely due to neglecting the contribution of electrons to the line broadening in the optically {\itshape thick} flare plasma.

\section{Spectral Line Emission}

To simplify the notation, we have in mind \textit{the Balmer series}. The intensity of the emission in the spectral line frequencies, $I^{(k)}_\nu$,
is equal to
\begin{equation}\label{L_01}
I^{(k)}_\nu=S^{(k)}_\nu\cdot\left[1-e^{-\tau^{(k)}_{\nu}-\tau_\nu}\right],
\end{equation}
where index $k$ identifies the upper level of the transition $k{\to}2$. We have denoted by $\tau^{(k)}_\nu$ the optical depth in the frequencies
of the $2{\to}k$ transition; the negative absorption being taken  into account:
\begin{equation}\label{L_02}
\tau^{(k)}_\nu=\varkappa^{(k)}_0\cdot{}a_\nu\cdot{}\mathcal{L}.
\end{equation}
Here, 
\begin{equation}\label{L_03}
\varkappa^{(k)}_0=%
\dfrac{\pi{}e^2}{m_ec}\,f_{2k}\cdot{}N_2%
\left[1-\dfrac{b_k}{b_2}\exp\left(-\dfrac{E_{k2}}{T_e}\right)\right],
\end{equation}
The spectral line intensity, $I_{k}$, is equal to the integral of \eqref{L_01}  over the line profile:
\begin{equation}\label{L_06}
I_{k}=\int{}S^{(k)}_\nu\cdot%
\left[1-e^{-\tau^{(k)}_\nu-\tau_\nu}\right]\,d\nu.
\end{equation}
In this section, we consider that the continuum absorption is negligible within the profile of a spectral line. So, we write the expression for the
source function taking into account only the line emission and absorption:
\begin{equation}\label{L_07}
S^{(k)}_\nu=B_\nu(T_e)\cdot%
\dfrac{1-e^{-y}}{{b_2}/{b_k}-e^{-y}}.
\end{equation}
The Planck function varies very slowly within the line profile. Therefore, the source function at the frequency of the line center, $S_k$, can be
taken  outside the integral sign:
\begin{equation}\label{L_06a}
I_{k}\approx{}S_k\cdot\int\limits_{\nu_l}^{\nu_u}%
\left[1-e^{-\tau_{\nu}(\nu_k)}e^{-\tau^{(k)}_\nu}\right]\,d\nu.
\end{equation}
For the same reason, we consider that the continuum optical depth  within the line profile is equal to $\tau_{\nu}(\nu_k)$.
 The limits of integration are the frequencies where the line emission merges into continuum. They are defined by the following
condition:
\begin{equation}\label{L_06b}
\tau^{(k)}_\nu(\nu_l)=\tau^{(k)}_\nu(\nu_u)=\tau_\nu(\nu_k).
\end{equation}
The simplified formulae \eqref{L_06a} and \eqref{L_06b} do not introduce significant inaccuracies in $I_k$, as the transition region is quite narrow.
Moreover, the emission intensity in this region is much lower than the intensity in the line core.

Note that there are some factors simplifying the calculation of integral \eqref{L_06a}. If the condition \eqref{pr_03} is satisfied, then the central
optical depth, $\tau_k$, is rather large. Consequently, the intensity in the line core frequencies coincides with a source function.

In those parts of the line profile where $\tau_{k}$ is not too large, we use for $a_\nu$ a simple approximation (\ref{Griem}), assuming
that for the frequency displacement up to $\beta_0$, the layer remains opaque. We introduce frequency displacement $\beta_0\geqslant3$ which
satisfies the condition
\begin{equation}\label{L_12}
\tau^{(k)}_\nu>5.
\end{equation}
In the offset range
\begin{equation}
0{\leq}|\beta|{\leq}|\beta_0|,
\end{equation}
it is assumed that the intensity is equal to the source function, and in the line wings we calculate the integral
\begin{equation}\label{L_13}
D_k=2\int_{\beta_0}^{\beta_u}%
\left(1-e^{-\tau_{\nu}(\nu_k)}\cdot{}%
e^{-\tau_k\beta^{-5/2}}\right)\,d\beta.
\end{equation}
Here, parameter $\tau_k$ is defined as
\begin{equation}\label{L_14}
\tau_k=\varkappa^{(k)}_0\cdot\dfrac32\cdot%
\cfrac{2\pi}{\mathcal{B}_{k2}\mathcal{E}_{0}}\cdot\mathcal{L},
\end{equation}
and the upper limit is calculated according to \eqref{L_06b}. Thus, the total energy emitted in the line is
\begin{equation}\label{L_15}
I_k=S_k\cdot\left(2\beta_0+D_k\right)\cdot%
\cfrac{\mathcal{B}_{k2}\mathcal{E}_{0}}{{2\pi}}\,,
\end{equation}
where the Stark width, $\mathcal{B}_{k2}\mathcal{E}_{0}$, is determined by formula  (\ref{Stark_eff}).

\section{Results}
We performed calculations of the spectral line and continuum intensity for different values of density and temperature of a homogeneous gas layer
under the conditions of flares on red dwarf stars. The layer thickness $\mathcal{L}$ is fixed. We calculated the optical depth at the center of the
Balmer series lines (see table \ref{table 1}) and the Menzel factors; in all cases the values of $b_k$  do not differ from unity (see table
\ref{table 2}). It is important  that there is no external source of radiation ionizing the hydrogen atom from the excited levels.

Let us address the results of calculations of the emission
spectrum. Figure~1 shows how the emission lines are ``immersed''
in a continuum with increasing density of the gas. The horizontal
axis represents the wavelength $\lambda$ in {\AA}ngstr\"oms,
vertical --- the logarithm of the ratio $I_\nu/B_\nu$. Balmer
lines are indicated by vertical line segments, its height is
proportional to the line center intensity. In all the graphs, the
following values of the parameters are assumed: $T_e=1.2$~eV,
$T_{ai}=6$~eV, and $\mathcal{L}=20$~km. The intensity of the
emission at the centers of the Balmer series lines is very close
to the Planck function at the electron temperature
$B_\lambda(T_e)$. Therefore, all the spectral lines have the same
upper boundary corresponding to zero on the vertical
axis. As the rise of the continuum occurs, first H$_\alpha$
``immerses'', then H$_\beta$ and finally high  members of the
Balmer series. This is due to the fact that continuum absorption
coefficient increases toward longer wavelengths.

\begin{figure}\label{Fig1}
\includegraphics[width=8cm]{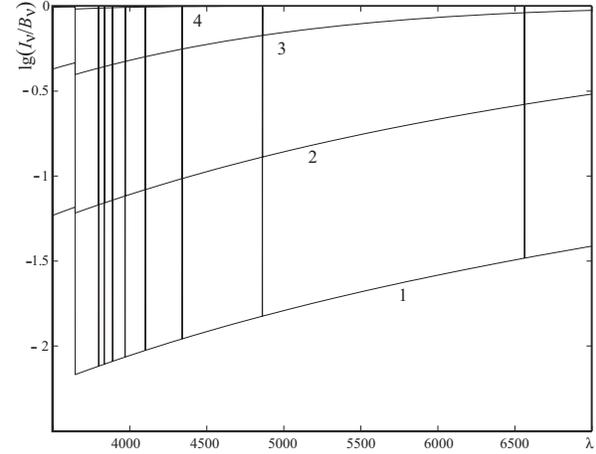}
\caption{The continuous spectrum and the height of the lines of the
Balmer series for the four density values. From bottom up:
$1{\cdot}10^{15}\text{ cm}^{-3}$, $3{\cdot}10^{15}\text{ cm}^{-3}$,
$1{\cdot}10^{16}\text{ cm}^{-3}$, $3{\cdot}10^{16}\text{ cm}^{-3}$.
Under each curve its number is written. The upper curve almost
coincides with the Planck function; the continuum absorption is not
negligible in this case.}
\end{figure}

We temporarily exclude spectral lines in order to more conveniently
trace how the continuous spectrum depends on the layer parameters.
Figure~2 represents the ratio $I_{\lambda}/B_{\lambda}$ for two
values of the electron temperature and three values of density.
Atomic--ion temperature is equal here to 10~eV, and the thickness of
the layer --- to 10~km. The values of the optical depth in a
continuous spectrum at the wavelength 4170 {\AA} are shown on each
curve. It is seen that increasing of the  continuum  opacity
essentially depends on the electron temperature: at higher values of
$T_e$, the transition to the Planck curve occurs at lower values of
$N_{\mathrm{H}}$. This fact is caused by increasing of the
occupation of excited levels ($k\geq3$) with increasing the electron
temperature. Indeed, the value of
$\varkappa^{(i)}_{\nu{k}}\propto{}N_k$, as it follows from equations
(\ref{s_03}) and (\ref{s_05b}).

While the gas density is increased, the intensity of the continuous spectrum $I_\lambda$
 tends
to the Planck function, remaining less than it. The curve portions
that are closer to the Planck curve are located in the red region of
the spectrum and at the blue side of the Balmer jump. When the
density  increases further, the blue part of the spectrum  (redder
at the Balmer jump) approaches the Planck function.

 The shape of the line core is largely determined by the ion-atom temperature, as the Doppler width, $\Delta\omega_{kn'}^D$, depends only on
 the square root of ${T_{ai}}$ (if the principal quantum numbers $k$ and $n'$ are unchanged). It is true that at higher values of $T_{ai}$,
 the approximation for the core by a gaussian curve  is valid for a higher number of members of the Balmer series.
\begin{figure}
\includegraphics[width=8cm]{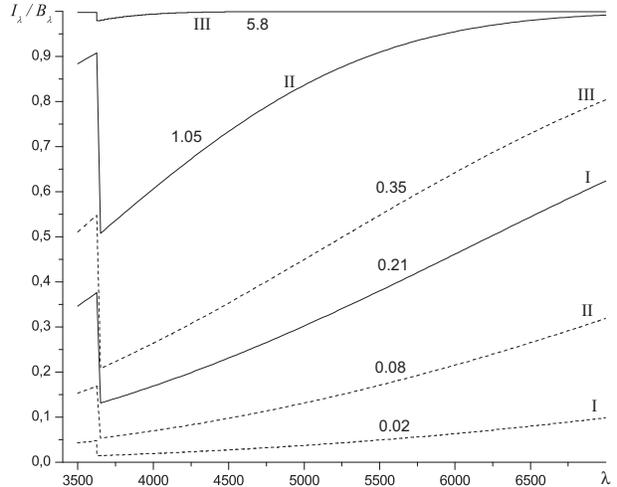}
 \caption{The continuous spectrum normalized to the Planck function (spectral lines are excluded). The dashed lines correspond to
$T_e=0.8$~eV, solid ones --- to $T_e=1.0$~eV, Numbers I, II, and III correspond to the following values of the total number densities of the hydrogen
particles: $2\cdot10^{15}$~cm$^{-3}$, $7\cdot10^{15}$~cm$^{-3}$, and $3\cdot10^{16}$~cm$^{-3}$, respectively. Next, the values of the optical depth
at wavelength $\lambda=4170$~{\AA} are shown for each curve. }
\end{figure}

\begin{table} \small \caption{Decimal logarithms of the optical depth at the centers of the Balmer lines ($T_e=1.2$~eV,
$T_{ai}=6$~eV, and $\mathcal{L}=20$~km)}\label{table 1}
\begin{tabular}{ccccc}
\hline $\log{}N$ & $\mathrm{H_\alpha}$ & $\mathrm{H_\beta}$ & $\mathrm{H_\gamma}$ & $\mathrm{H_\delta}$\\
15.0& 6.40& 5.54 &5.06 &4.73 \\
15.5& 6.90 & 6.04 &5.56 &5.23 \\
16.0&7.40 &6.54 &6.07 &5.74 \\
16.5&7.90 &7.04 & 6.57&6.24 \\
\hline
\end{tabular}
\end{table}

\begin{table} \small \caption{Menzel factors for the discrete levels of the hydrogen atom}\label{table 2}
\begin{tabular}{cccccc}
\hline $\log{}N$ &1 &2& 3 & 4 &5\\
15.0& 1.030 &1.031 &1.006 &1.001&1.000 \\
15.5& 1.009&1.009 &1.003 &1.002 &1.001\\
16.0&1.003&1.003 &1.002 &1.001 &1.000\\
16.5&1.001 &1.001 &1.001 &1.001&1.000\\
\hline
\end{tabular}
\end{table}

\section{Discussion}

The main result of our work is the calculated plasma emission spectrum in a broad interval of the continuum optical depths, from the transparent gas
to the gas whose emission intensity is close to the Planck function. The proximity to the Planck function in the case of strong self-absorption is
due to the mechanism of impact ionization and excitation (from electron collisions) of the plasma, whereas radiative processes are secondary. The
approach used in this paper fits well for red dwarf stars with their effective temperature of about 3000~K. However, the case of G- and F-type dwarfs
requires taking into account the photospheric emission, e.g., photoionization of hydrogen atoms from exited levels.  In the case of appreciable
emission and absorption in the continuum its (continuum) influence must be taken into account in the calculation of spectral line intensities. Here
our work develops the approach used in \cite{dra80} and \cite{Katsova_91}, where the line intensities were calculated under the assumption of the
transparent continuum bringing small contribution to the emission.

In the framework of the considered problem the intensity of the subordinate lines is close to Planck for any layer parameters implemented during the
flares on red dwarf stars. Therefore, the relative height of the lines depends on the optical depth in the continuum. If it is small then strong
lines are observed against the background of faint continuum, which is described by formulae valid for transparent gas. In the case of appreciable
self-absorption in the continuum its spectrum becomes close to Planck; thereby, the lines weaken. However, in the homogeneous layer model no strong
lines on the background of the Planck continuum observed by \cite{Kowalski13} are obtained. Probably, we see in the flares the total emission from
several layers; some of them are transparent in the continuum, while other ones are not. We hope to get a better agreement with observations by
calculations of the radiative cooling behind the shock front under the conditions of the red dwarf chromospheres, since the shocked gas can include
at the same time equilibrium and non-equilibrium regions.

\section*{Acknowledgments}

E.S. Morchenko is grateful to Prof. Yu.K. Zemtsov for valuable remarks. This work was partly supported by the Russian Foundation for Basic Research
(project code 14-02-00922), and the Scientific School (project code 1675.2014.2 NSh). We thank the anonymous referee for very useful suggestions and
comments.


\begin{thebibliography}{}
\bibitem[Gershberg \& Pikel'ner(1972)]{Gersh_72} Gershberg, R.E., Pikel'ner S.B.: Comments Astrophys. Space Phys. {\bfseries4},  113 (1972).
\bibitem[Gershberg(2005)]{Gersh_2005} Gershberg, R.E.: Solar-Type Activity in Main-Sequence Stars, (Springer) (2005).
\bibitem[Kostyuk \& Pikel'ner(1975)]{Pik_74} Kostyuk, N.D., Pikel'ner, S.B.: \sovast{} {\bfseries18}, 590 (1975).
\bibitem[Katsova et al.(1981)]{liv80} Katsova, M.M., Kosovichev, A.G., Livshits, M.A.: Astrophysics. {\bfseries 17}, 156
(1981). doi: 10.1007/BF01005196
\bibitem[Fadeyev \& Gillet(2001)]{FG2001} Fadeyev, Yu.A., Gillet, D.: \aap{} {\bfseries 368}, 901 (2001). arXiv:astro-ph/0603510. doi:10.1086/506268
\bibitem[Belova et al.(2014)]{BBMN2014} Belova, O.M., Bychkov, K.V., Morchenko, E.S., Nizamov, B.A.: Astron. Reports. {\bfseries 58}, 650 (2014)
doi:10.1134/
S1063772914090029
\bibitem[Kowalski et al.(2013)]{Kowalski13} Kowalski, A.F., Hawley, S.L., Wisniewski, J.P., Osten, R.A., Hilton, E.J., Holtzman, J.A., Schmidt, S.J., Davenport, J.R.A.:
\apjs{} {\bfseries 207}, 57 (2013). 1307.2099. doi:10.1088/0067-0049/207/1/15
\bibitem[Drake \& Ulrich(1980)]{dra80} Drake, S.A., Ulrich, R.K.: \apjs{} {\bfseries 42}, 351, 1980.
\bibitem[Zel'dovich \& Raizer (1967)]{Pick-1961} Zel'dovich, Ya.B., Raizer, Yu.P.: Physics of Shock Waves and High-Temperature Hydrodynamic Phenomena,
(Academic Press) (1967).
\bibitem[Johnson(1972)]{Johnson72} Johnson, L. C.: \apj{} {\bfseries 174}, 227 (1972).
\bibitem[Sobel'man(1963)]{sob63} Sobel'man, I.I.: Introduction to the Theory of Atomic Spectra, (Pergamon Press) (1972).
\bibitem[Vidal(1966)]{vid66} Vidal, C.R.: J. Quant. Spectrosc. Rad. Transfer {\bfseries 6}, 461 (1966).
\bibitem[Holstein(1947)]{hol47} Holstein, T.: Phys. Rev. {\bfseries{72}}, 1212 (1947).
\bibitem[Biberman et al. (1982)]{bib82} Biberman, L.M., Vorob'ev, V.S., Yakubov, I.T.: Kinetics of Nonequilibrium
Low-Temperature Plasmas, (Plenum Publishers) (1987).
\bibitem[Smirnov(1972)]{smi72} Smirnov, B.M.: Physics of weakly ionized gas, (Nauka) (1972).
\bibitem[Sobolev \& Grinin(1996)]{sob96} Sobolev, V.V., Grinin V.P.: In: Pallavicini, R., Dupree, A.K. (eds.) Cool Stars, Stellar Systems, and the Sun,
9th Cambridge Workshop ASP Conference Series. {\bfseries{109}}, 629 (1996).
\bibitem[Sobolev \& Grinin(1995)]{sob95} Sobolev, V.V., Grinin, V.P.: Astrophysics. {\bfseries 38}, 15 (1995). doi:10.1007/BF02113956
\bibitem[Stehle(1994)]{ste94} Stehle, C.: \aaps{} {\bfseries 104}, 509 (1994).
\bibitem[Ivanov(1969)]{iva69} Ivanov, V.V.: Transfer of Radiation in Spectral Lines, (U.S. Government Printing Office) (1973).
\bibitem[Mihalas(1974)]{mih78} Mihalas, D.: Stellar Atmospheres, (W.H. Freeman and Co.) (1974).
\bibitem[Eason et al. (1992)]{eas92} Eason, E.L.E., Giampapa, M.S., Radick, R.R., Worden, S.P., Hege, E.K.: \aj{} {\bfseries104}, 1161 (1992).
\bibitem[Underhill \& Waddell (1959)]{und59} Underhill, A.B., Waddell, J.H.: National Bureau Standards Circular No. 603 (1959).
\bibitem[Katsova et al. (1991)]{Katsova_91} Katsova, M.M., Livshits, M.A., Butler, C.J., Doyle, J.G.: \mnras{} {\bfseries 250}, 402 (1991).
\end{thebibliography}
\end{document}